\documentclass{article}
\usepackage[utf8]{inputenc}
\usepackage{authblk}
\usepackage{amsmath}
\usepackage{amssymb}
\usepackage[numbers]{natbib}
\usepackage[colorlinks=true,linkcolor=blue,unicode]{hyperref}
\usepackage{graphicx}
\usepackage{subfig}

\title{How to be an extremist}
\author[1]{Tomáš Fürst}
\author[1]{Anna Minarovičová}
\author[1, 2]{Františka Sandroni}
\author[3, *]{Jakub Dostál}

\affil[1]{Palacký University, Olomouc}
\affil[2]{University of Florence}
\affil[3]{Abradatas s.r.o.}
\affil[*]{Corresponding author: Jakub Dostál, dost.jak@gmail.com}

\begin{document}

\maketitle

\begin{abstract}
We present a toy model of opinion spreading in a society
which combines a self-reinforcing mechanism with diffusion.
The relative strength of these two mechanisms -- called the affectability of the system --
is a free parameter of the model.
The model is run on a scale-free network and its asymptotic behaviour is investigated.
A surprising emergent effect is observed:
If every individual becomes more attentive to the opinions of others,
the society as a whole switches from a plurality of nuanced opinions
into the state of an absolute consensus on an extreme opinion. 
This counter-intuitive emergent behaviour may help to explain certain paradoxes
in human history.
\end{abstract}    

\section{Introduction}
\noindent
We have always wondered why people have the opinions they have.
Why did mediaeval Europeans think their souls would pass through purgatory after they died~\cite{Zaleski2018}?
Why did the Germans before the war think that Jews posed a mortal threat to their nation?~\cite{Weitz2015}
Why do some people today think that passenger aircraft spray us with poisonous chemicals~\cite{chemtrails}?
Where do these spontaneous clusters of consensus originate?
From a distance of a few generations, a few hundred kilometers, or a few social classes they look almost incredible.
Nevertheless, they are real and sometimes they matter more than governments, parliaments, and constitutions~\cite{Ferguson2018}.

\section{The Model}
Let me tell you how I come to an opinion.
I think today what I thought yesterday, just a little bit more.
That is if I haven't talked about the matter with anyone and if I haven't received any new information.
Without any interaction with the world, my opinions tend to strengthen themselves.
(My wife says only men work this way, and that is the problem.)

\begin{figure}[ht!]
\includegraphics[width=\textwidth]{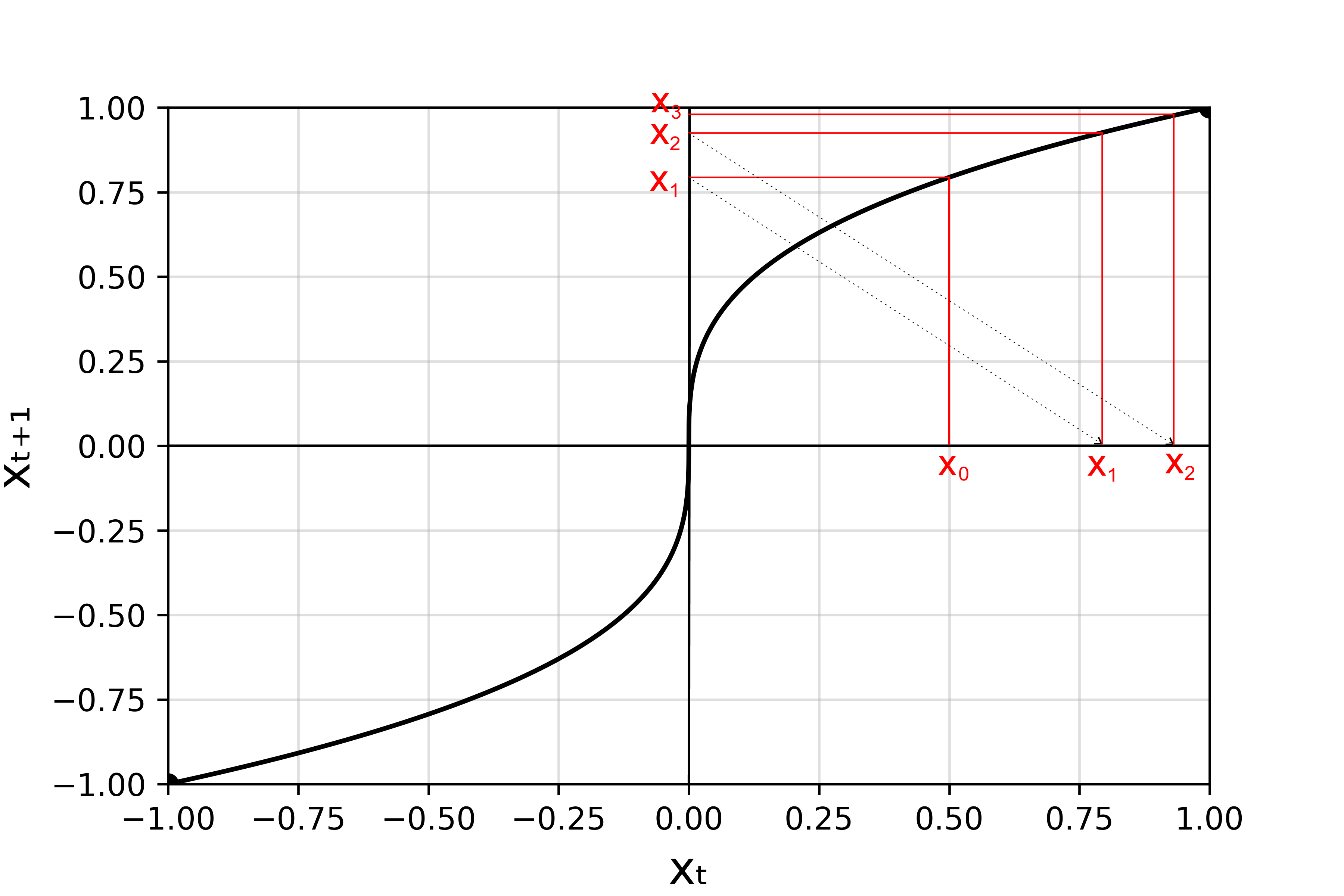}
\caption{Self-amplification of an opinion. 
An opinion $x_t$ of a person is modelled by a number between 0 and 1.
In a hypothetical world where no one interacts with anyone else and no one receives any new information,
the opinion amplifies itself through equation (\ref{tvrdohlav}). 
The figure shows the graph of $f(x) = \sqrt[3]{x}.$
Three iterations of this mapping are shown for $x_0 = 0.5.$}
\label{fig:tvrdohlav}
\end{figure}

Let us assume that my opinion on a certain matter (e.g. \textit{Brexit}) lies somewhere between 
``absolutely nay'' ($x=-1$) and ``aye by all means'' ($x=+1$). 
The self-amplification of the opinion may be modelled by the relation
\begin{equation}\label{tvrdohlav}
	x_{t+1} = \sqrt[3]{x_t},
\end{equation}
where $x_t$ denotes my opinion yesterday and $x_{t+1}$ my opinion today.
The choice of the cube root is not essential;
any odd extension of an increasing concave function $f:[0,1] \to [0,1]$ would work.
Figure \ref{fig:tvrdohlav} shows the behaviour of a single opinion repeatedly 
amplified through function (\ref{tvrdohlav}).
In a hypothetical world where no one interacts with anyone else and no one receives any new information, 
all opinions slowly drift to an extreme. Those who started with a slightly positive opinion $(x_0>0),$ 
will end up as extremists $(x_\infty=1).$ Those who started with a slightly negative opinion $(x_0<0)$
will also end up extreme $(x_\infty = -1).$ Only the phlegmatic $(x_0=0)$ will stay phlegmatic forever $(x_\infty=0).$

In the real world, however, we are inter-connected by a web of friendship, hostility, media, and social networks
where we exchange information and opinions. Whenever I read something about \textit{Brexit} in the newspaper,
see something on TV, or talk to friends or foes, I correct my opinion according to what I learn.
This mechanism may be captured by the relation
\begin{equation}\label{difuse}
	x_{t+1} = \frac{ y_t^1 + y_t^2 + \dots + y_t^N}{N}
\end{equation}
where $x_{t+1}$ denotes my opinion today and $y_t^i$ denotes the opinion of my $i-$th source yesterday.
This relation describes a completely affectable person whose opinion today is a simple arithmetic mean of the opinions of his friends yesterday
-- the dream of a marketing specialist come true.

A more realistic mechanism for the spreading of opinions in a society may be a combination of the two approaches described above. Most classical approaches do not include opinion drift (\ref{tvrdohlav}) 
(see ~\cite{Noorazar2019} for a recent review);
however, we will show that it is a key factor which makes the model non-linear and thus interesting.

Let us assume a group of $N$ individuals (vertices/nodes) inter-connected by certain links (edges).
The links are described by the adjacency matrix $A,$ the elements of which are either zeroes or ones,
\begin{equation}
    a_{ij} =
    \begin{cases}
        $1$,  & \text{if nodes}\ i\ \text{and}\ j\ \text{are connected by an edge,}\\
        $0$,  & \text{if nodes}\ i\ \text{and}\ j\ \text{are not connected}.
    \end{cases}
\end{equation}
For the sake of simplicity, we assume that the graph is unweighted 
(all the links are positive with a unit weight)
and undirectional $(a_{ij} = a_{ji}).$
It is straightforward to include directional edges (The Economist influences me and not vice versa),
more or less important edges (my wife has more influence on me than my cat),
and edges with negative weights (it is a matter of honour not to align my opinion with Russia Today).

Let us focus on a certain matter that may be subject to an opinion.
The value $x_t^i \in [-1,1]$ denotes the opinion of individual $i$ at time $t.$
The combination of the mechanisms (\ref{tvrdohlav}) and (\ref{difuse}) reads
\begin{equation}\label{oba}
	x_{t+1}^i = \alpha \, \sqrt[3]{x_t^i} + \frac{1-\alpha}{k_i} \, \sum_{j=1}^N a_{ij} x_t^j
\end{equation}
The term $k_i$ denotes the degree of the respective vertex (i.e. the number of its neighbours),
which is given by
$$	k_i = \sum_{j=1}^N a_{ij}. $$
The term $(1-\alpha) \in [0,1]$ captures the ``affectability'' of the system.
For $\alpha$ close to one, mechanism (\ref{tvrdohlav}) prevails, while for $\alpha$ close to zero mechanism (\ref{difuse}) prevails.
For simplicity, let us assume that all individuals have the same affectability.

\section{Simulation Results}
The model is so simple that it contains only two features to investigate:
(1) the affectability $\alpha$, 
and (2) the type of connectivity of the network given by the adjacency matrix $A$. 
The behaviour of the model naturally also depends on the initial condition,
which will be sampled randomly from a uniform distribution on $[-1,1]$.

\begin{figure}[!tbp]
  \centering
  \subfloat[Initial state.]{\includegraphics[width=0.4\textwidth]{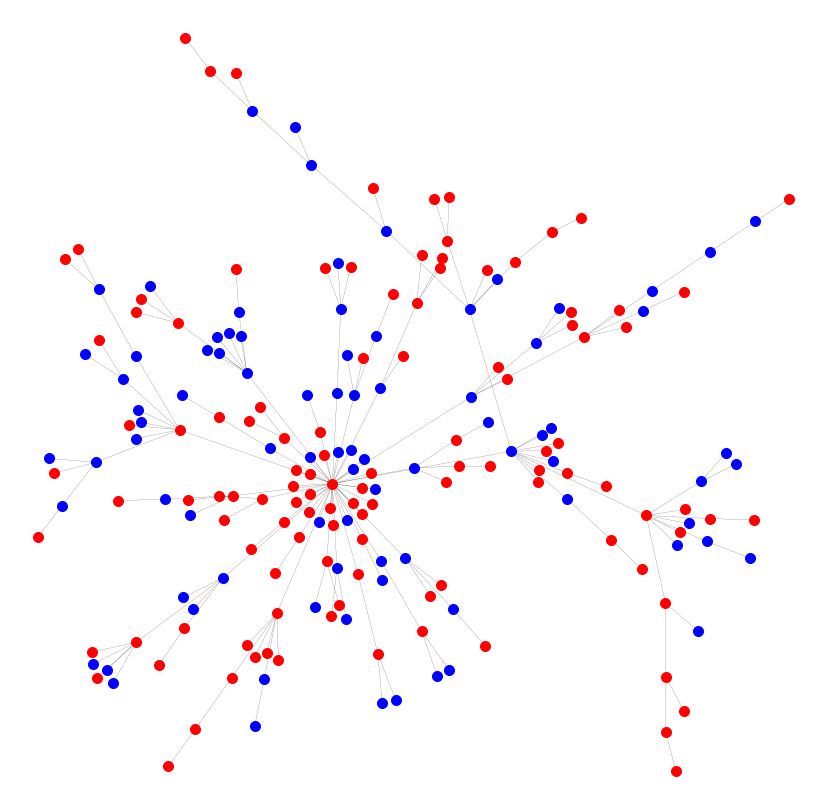}\label{fig:pre}}
  \hfill
  \subfloat[Final state.]{\includegraphics[width=0.4\textwidth]{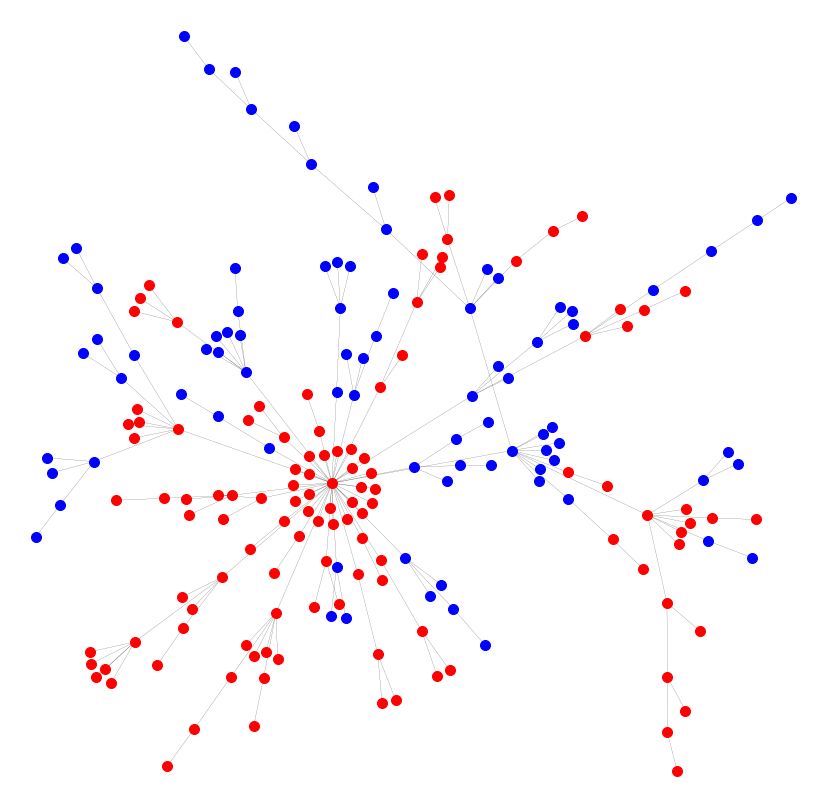}\label{fig:post}}
\caption{A typical scale-free network.
This is an illustrative example of a small network ($N=200$ and $m=2$);
a network that was five times larger was used for the simulations.
The initial opinion distribution was drawn from a uniform distribution.
Nodes with a positive opinion (blue) and nodes with a negative opinion (red).}
\label{fig:net}
\end{figure}

It has been shown~\cite{Barabasi1999} that the human interaction network can be captured
by what are called scale-free models.
Scale-free networks are characterized by a vast majority of vertices of 
low degree and a few vertices with a very large degree (hubs).
More precisely, the frequency of vertices of degree $k$ decreases as $k^{-\gamma},$
where $\gamma$ usually lies between 2 and 3.
We chose such a scale-free model (Barabási-Albert model~\cite{Barabasi1999}) 
for the simulation of opinion spreading for various values of  $\alpha.$
We generated a scale-free network by means of the NetworkX package~\cite{NetworkX}.
The network generation process started with a small network of $m < N$ nodes. 
Nodes were added until the desired network size was reached. 
The probability of the connection of a new node $A$ to an already existing node  $B$ 
was proportional to the degree of $B.$
See Figure \ref{fig:net} for an example of a simulated scale-free network.

\begin{figure}[ht!]
\includegraphics[width=\textwidth]{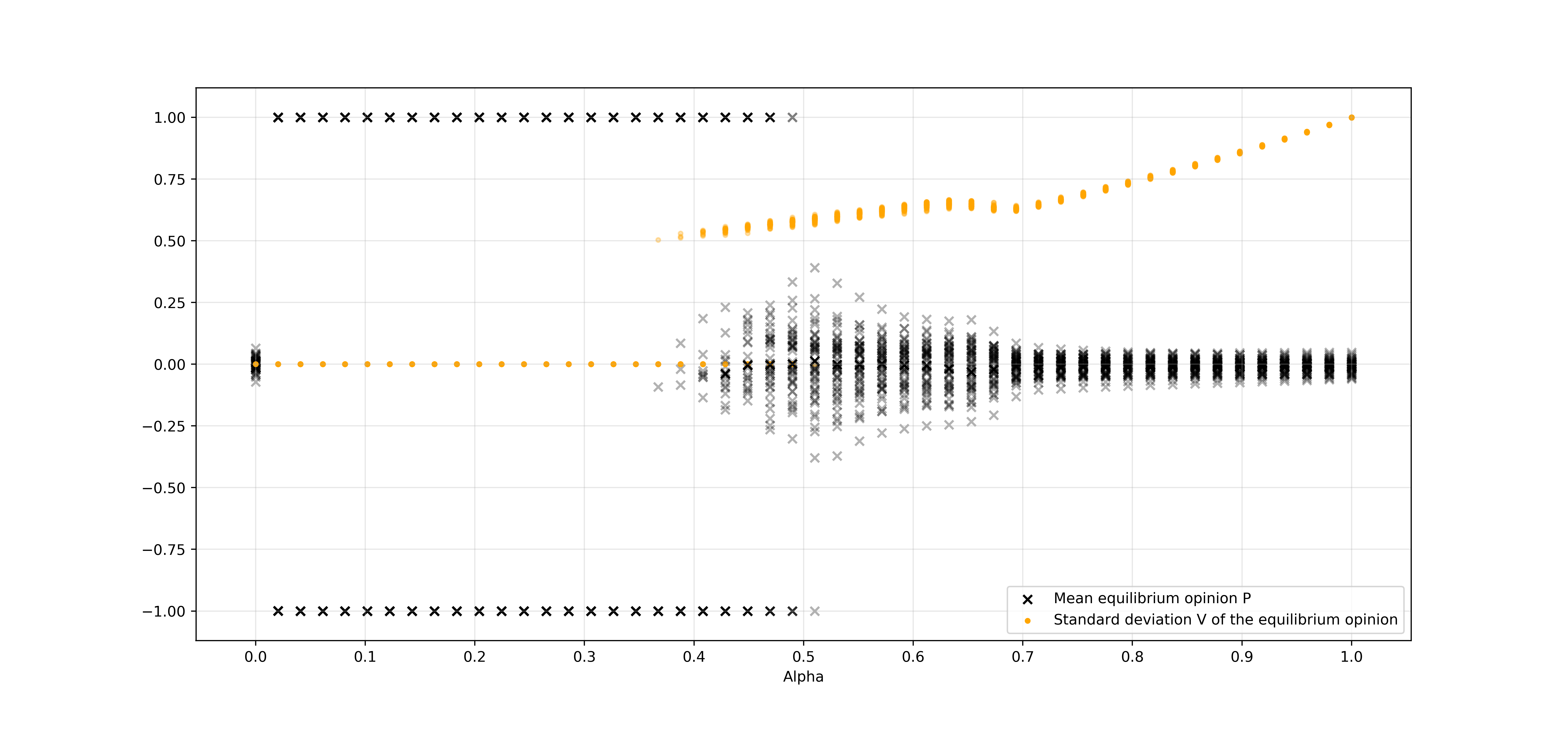}
\caption{
Summary of the 2,500 runs of the model on a scale-free network with $N=1000$ and $m=2.$
For each simulation, the mean equilibrium opinion $P$ is shown (black cross) and also
the standard deviation $V$ of the equilibrium opinion (orange dot).
The horizontal axis shows the affectability of the system.
See the text for details.}
\label{fig:ba-model}
\end{figure}

Let us now simulate the spreading of opinions
\footnote{The source code of the simulation is available at \url{https://github.com/DostalJ/OpinionModel}.} 
by means of equation (\ref{oba}) on a scale-free network
with the parameters $N=1000$ and $m=2$ (a network five times larger than the one in Figure \ref{fig:net}).
The value of $\alpha$ is increased from $0$ to $1$ with step $0.02.$
For each value of  $\alpha,$ 50 simulations of the model (\ref{oba}) are performed, 
each time with a different initial condition randomly sampled from a uniform distribution on $[-1,1].$
The simulation is terminated after an equilibrium has been reached.
In the final equilibrium state, the mean opinion is computed
$$	P = \frac{1}{N} \sum_{i=1}^N x_\infty^i $$
and its variability is measured by the standard deviation
$$	V = \frac{1}{N} \sqrt{\sum_{i=1}^N (x_\infty^i - P)^2} $$
For each simulation, the mean equilibrium opinion $P$ is depicted by a black cross 
in Figure \ref{fig:ba-model}.
Its standard deviation is plotted by an orange dot. 
Altogether, $50 \times 50$ simulations are performed, so Figure \ref{fig:ba-model} 
contains $2500$ crosses and the same number of dots.
For example, the simulation for $\alpha = 0.1$ yields all the crosses either at $P=1$ or $P=-1,$
with the standard deviation always at zero. 
This means that all the simulations ended with everyone taking the same extreme opinion, 
either $x=-1$ or $x=1$ with no variation.
This is an absolute consensus on an extreme -- 
everyone takes the same extreme opinion and the consensus is absolute.
However, looking e.g. at $\alpha = 0.8,$ a different picture appears.
The crosses showing the mean equilibrium opinion $P$ form a line segment between $P \sim -0.1$ and $P \sim 0.1,$
while the standard deviation stays around $V=0.6$ for all the simulation runs.
This means that the system always stabilized in a plurality of non-extreme opinions.
There may be individuals taking an extreme opinion $x=+1$ or $x=-1;$
however, most of the nodes are moderate.
The fraction of positive and negative opinions in the equilibrium is similar, thus $P \sim 0$.

The most remarkable region in Figure \ref{fig:ba-model} lies around $\alpha = 0.65,$
where there is a transition between the absolute consensus on an extreme and plurality of opinions.
A small change in $\alpha$ (e.g. from $0.67$ to $0.62$) may lead to the sudden 
emergence of an absolute consensus on an extreme
without any change in the structure of the network.

\section{Discussion}
If there is anything realistic about this toy model, 
we should wonder which processes may alter the value of $\alpha.$
Education is an obvious candidate.
Socrates' famous quote ``Scio me nihil scire'' (which probably comes neither from Socrates nor in this form)~\cite{Socrates-wiki}
reminds us that the more we know, the more nuanced the opinions we form are and the less convinced about them we are.
Let us assume that education makes one more attentive to the opinions of others
and thus \emph{increases} the affectability of the system (and consequently reduces $\alpha$).
This assumption is very natural -- much research shows a correlation between lower education and extreme opinions~\cite{Davies2003}.
Suddenly, we face a fascinating emergent behaviour of the system:
as the level of education in a society grows, affectability increases and the society may quite suddenly
switch from a healthy plurality of moderate opinions into the state of an absolute consensus on an extreme.
This transition is very counter-intuitive.
On the level of an individual, higher affectability makes one more attentive to the opinions of others.
However, on the level of the society, the effect of higher affectability is quite the opposite in a rather dramatic way.

Naturally, one may disagree with the assumptions of the toy model,
and doubt the interpretation of its parameters.
However, there is at least one reason to give it a second thought:
many historians keep wondering why people in pre-war Germany -- 
probably the most educated society in the world at that time~\cite{Elon2004} -- 
succumbed to mass psychosis and decided to exterminate many of their neighbours.
As this toy model shows, emergent effects in complex networks may be very counter-intuitive.

\medskip

\bibliographystyle{abbrvnat}
\bibliography{references}

\end{document}